\documentclass[aps,prx,superscriptaddress,twocolumn,floatfix,citeautoscript,longbibliography,hyperlinks]{revtex4-1}

\usepackage{graphicx}
\usepackage{dcolumn}
\usepackage{bm}
\usepackage[dvipsnames]{xcolor}
\usepackage[colorlinks=true,urlcolor=blue,citecolor=blue, linkcolor = blue]{hyperref}
\usepackage{amssymb,amsmath}
\newcommand{\bra}[1]{\left\langle #1\right|}
\newcommand{\ket}[1]{\left|#1\right\rangle}

\renewcommand{\vec}{\mathbf}
\makeatletter
\newsavebox{\@brx}
\newcommand{\llangle}[1][]{\savebox{\@brx}{\(\m@th{#1\langle}\)}%
  \mathopen{\copy\@brx\kern-0.5\wd\@brx\usebox{\@brx}}}
\newcommand{\rrangle}[1][]{\savebox{\@brx}{\(\m@th{#1\rangle}\)}%
  \mathclose{\copy\@brx\kern-0.5\wd\@brx\usebox{\@brx}}}
\makeatother

\begin{document}

\title{Quantum walks on random lattices: Diffusion, localization and the absence of parametric quantum speed-up}

\author{Rostislav Duda}
\affiliation{Computational Physics Laboratory, Physics Unit, Faculty of Engineering and
Natural Sciences, Tampere University, FI-33014 Tampere, Finland}
\affiliation{Helsinki Institute of Physics, FI-00014 University of Helsinki, Finland}
\affiliation{Department of Applied Physics, Aalto University, FI-00076 Aalto, Finland}
\author{Moein N. Ivaki}
\affiliation{Computational Physics Laboratory, Physics Unit, Faculty of Engineering and
Natural Sciences, Tampere University, FI-33014 Tampere, Finland}
\affiliation{Helsinki Institute of Physics, FI-00014 University of Helsinki, Finland}
\author{Isac Sahlberg}
\affiliation{Computational Physics Laboratory, Physics Unit, Faculty of Engineering and
Natural Sciences, Tampere University, FI-33014 Tampere, Finland}
\affiliation{Helsinki Institute of Physics, FI-00014 University of Helsinki, Finland}
\author{Kim P\"oyh\"onen}
\affiliation{Computational Physics Laboratory, Physics Unit, Faculty of Engineering and
Natural Sciences, Tampere University, FI-33014 Tampere, Finland}
\affiliation{Helsinki Institute of Physics, FI-00014 University of Helsinki, Finland}
\author{Teemu Ojanen} \email{Email: teemu.ojanen@tuni.fi}
\affiliation{Computational Physics Laboratory, Physics Unit, Faculty of Engineering and
Natural Sciences, Tampere University, FI-33014 Tampere, Finland}
\affiliation{Helsinki Institute of Physics, FI-00014 University of Helsinki, Finland}

\begin{abstract}
Discrete-time quantum walks, quantum generalizations of classical random walks, provide a framework for quantum information processing, quantum algorithms and quantum simulation of condensed matter systems. The key property of quantum walks, which lies at the heart of their quantum information applications, is the possibility for a parametric quantum speed-up in propagation compared to classical random walks. In this work we study propagation of quantum walks on percolation-generated two-dimensional random lattices. In large-scale simulations of topological and trivial split-step walks, we identify distinct pre-diffusive and diffusive behaviors at different time scales. Importantly, we show that even arbitrarily weak concentrations of randomly removed lattice sites give rise to a complete breakdown of the superdiffusive quantum speed-up, reducing the motion to ordinary diffusion. By increasing the randomness, quantum walks eventually stop spreading due to Anderson localization. Near the localization threshold, we find that the quantum walks become subdiffusive. The fragility of quantum speed-up implies dramatic limitations for quantum information applications of quantum walks on random geometries and graphs.           

\end{abstract}

\maketitle

\section{Introduction} 

The concept of a random walk occupies a central role in physics, mathematics, statistics and information processing. In the looming quantum information era, it is no surprise that quantum-mechanical generalizations of the random walk, quantum walks, have become a subject of broad interest for physicists with various specializations~\cite{kempeoverview,PhysRevA.48.1687,PhysRevA.58.915,PhysRevB.86.195414,aharonov2001quantum,karski2009quantum,PhysRevX.3.031005}. Combining ideas from quantum information processing to condensed matter physics and beyond, discrete-time quantum walks (DTQW) have been proposed as a general framework for a variety of quantum algorithms and studying condensed-matter phenomena~\cite{PhysRevA.67.052307,PhysRevLett.121.100502,PhysRevLett.102.180501,gong2021quantum,kendon2006random,mallick2019simulating,PhysRevLett.124.050502,chawla2020discrete,PhysRevLett.103.090504,PhysRevLett.121.100501,PhysRevLett.129.046401}. The wide range of possible applications of quantum walks include quantum computing, quantum cryptography, quantum search algorithms on the quantum information side and a simulation of fundamental properties of quantum phases of matter on the condensed-matter side, some of which have already seen experimental realization~\cite{schreiber20122d,PhysRevA.70.042312,PhysRevX.7.031023,PhysRevLett.122.020501,cardano2016statistical,xiao2017observation,PhysRevLett.128.120401,PhysRevA.72.062317}.

\begin{figure}[ht]
    \centering
    \includegraphics[width=0.8\columnwidth]{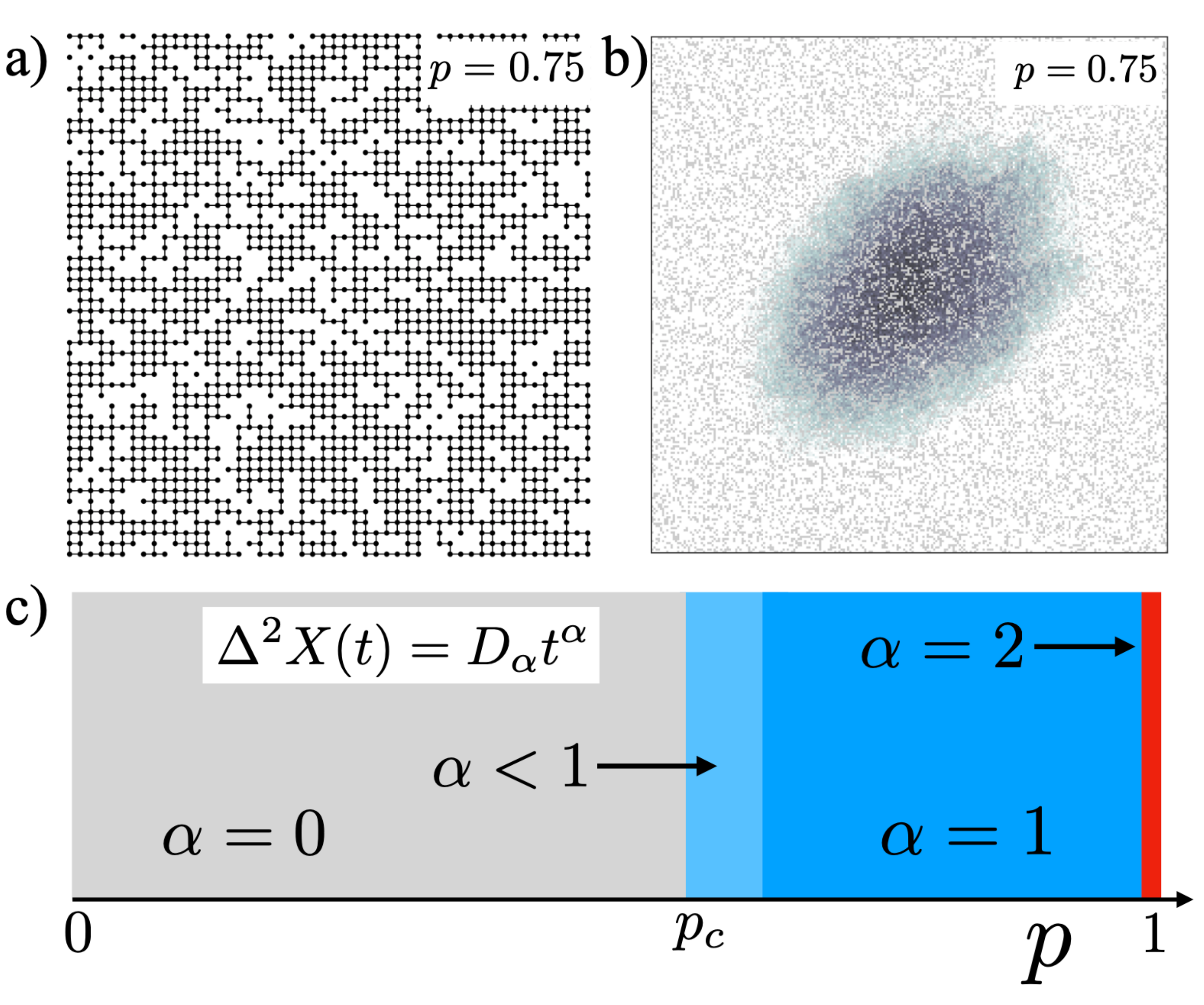}
    \caption{Quantum walks in random geometry. a): Single realization of percolation lattices where each site of a square lattice is present with probability $p$ (figure corresponds to $p=0.75$). b): Probability distribution of a quantum walk which started at the origin at $t=0$ and propagated $t=256$ steps on a single realization of a $p=0.75$ lattice. Color scale logarithmic to improve visibility. c): Phase diagram of the asymptotic long-time motion of quantum walks on the studied lattices. The three regions, in the order of decreasing randomness, correspond to localized ($\alpha=0$), subdiffusive ($\alpha<1$) and diffusive ($\alpha=1$) phase. Superdiffusive behavior is restricted to a pristine square lattice with $p=1$, where motion is ballistic.    }
    \label{fig:1}
\end{figure}

The main attractive feature of quantum walks is the possibility of a parametric quantum speed-up of propagation relative to a classical random walk~\cite{childs2003exponential,PhysRevA.65.032310,PhysRevLett.123.140501,PhysRevLett.104.050502}. While random walks in general give rise to a diffusive motion, characterized by a mean square displacement (MSD) $\Delta^2 X\propto t$ which grows linearly in the number of time steps $t$, quantum walks on regular lattices and graphs may reach a quadratic speed-up $\Delta^2 X\propto t^2$ corresponding to a ballistic spreading. The quadratic quantum speed-up can be harnessed, for example, to realize a version of the celebrated Grover's search algorithm~\cite{grover1996fast,PhysRevLett.79.325} with a square root reduction in the execution time compared to classical algorithms~\cite{portugal2013quantum,PhysRevA.70.022314,ambainis2004coins}. Thus, it is a matter of central importance for applications of quantum walks to understand under what circumstances quantum walks can maintain a speed-up compared to the standard diffusion. In particular, \textit{can parametric quantum speed-up persist in irregular or random structures}? If so, quantum-walk-based algorithms could be employed, for example, to speed up element search in large irregular and random structures. 

In this work, we study the propagation of quantum walks in random lattice geometries through the lens of condensed matter physics. We consider a family of DTQWs~\cite{kitagawa2012topological,PhysRevA.82.033429} on infinite random lattices generated by site percolation, as illustrated in Figs.~\ref{fig:1} a) and b). Motivated by the fact that topological states of matter are generically more robust to random perturbations, we implement split-step quantum walks with a tuneable topological invariant and explore the spreading of topological and trivial quantum walks. By carrying out large-scale numerical simulations up to $t=10^4$ time steps, we find that pre-diffusive and transient kinetics dominate at intermediate time scales $t_{\text{decay}}$ determined by the degree of randomness. In the long-time limit, we observe that the configuration-averaged MSD follows the generalized diffusion Ansatz $\Delta^2 X\propto t^\alpha$, and extract the exponent $\alpha$. Our findings are summarized in Fig.~\ref{fig:1} c), which indicates that even weak randomness will give rise to the breakdown of the superdiffusive quantum speed-up. 
With increasing random dilution, the quantum walks will eventually halt due to Anderson lozalization at the critical density $p_c$. Moreover, in the vicinity of $p_c$, the system becomes subdiffusive. As discussed below, the absence of superdiffusion implies severe limitations for obtaining quantum speed-up in quantum-walk-based applications on random lattices and graphs.

\section{Quantum walks on random lattices}

\subsection{Topological walks on regular lattice}

Before discussing random geometries, we first explore the properties of the studied topological split-steps walk on a regular lattice. In general, a DTQW on a square lattice is defined for a point-like walker with an internal $n$-level degree of freedom referred to as a quantum coin. The quantum state of a walker belongs to a Hilbert space $\mathcal{H}=\mathbb{Z}^{2} \otimes \mathbb{C}^{n}$ with a basis $ \ket{x, y} \otimes \ket{\mathbf{s}}$, where $\ket{x, y}$ refers to the position states and $\ket{\mathbf{s}}$ to the internal coin states. In analogy to a random walk, a quantum walk is defined as a sequence of quantum coin operations and conditional translations depending on the coin state. A walker is initially located at the origin, and a single step of a walk is generated by a unitary $\hat{U}$ which propagates the walker state as $\ket{\psi(t+1)}=\hat{U}\ket{\psi(t)}$. Hereinafter we will assume a spin-$\frac{1}{2}$ quantum walk with $n = 2$. To define the unitary, one needs a translation operator 
\begin{align}
        \hat{T}(\vec{\delta}) = \sum_{\vec{r} \in \mathbb{Z}^{2}}\Big[ \ket{\uparrow}\bra{\uparrow} \otimes  \ket{\vec{r}+\vec{\delta}}\bra{\vec{r}} +\nonumber  \ket{\downarrow}\bra{\downarrow} \otimes  \ket{\vec{r}-\vec{\delta}}\bra{\vec{r}} \Big],
\end{align}
which shifts the position of a walker by the vector $\pm\vec{\delta}$ depending on the coin state, and a coin operator 
\begin{align}
  \hat{R}(\theta) = e^{-\frac{i\theta}{2}\hat{\sigma}_{y}}\nonumber.
\end{align}

Following Ref.~\cite{kitagawa2012topological}, by introducing the primitive shift vectors $\delta_1=(1,1)$, $\delta_2=(0,1)$ and $\delta_3=(1,0)$, we define a split-step unitary as
\begin{align}
    \hat{U}_{\mathrm{2D}}(\theta_{1},\theta_{2}) = \hat{T}(\vec{\delta}_3)  \hat{R}(\theta_{1}) \hat{T}(\vec{\delta}_2) \hat{R}(\theta_{2})  \hat{T}(\vec{\delta}_1)  \hat{R}(\theta_{1}).
\label{eq:dtqw_2d}
\end{align}

\begin{figure}[h]
    \centering
    \includegraphics[width=0.85\columnwidth]{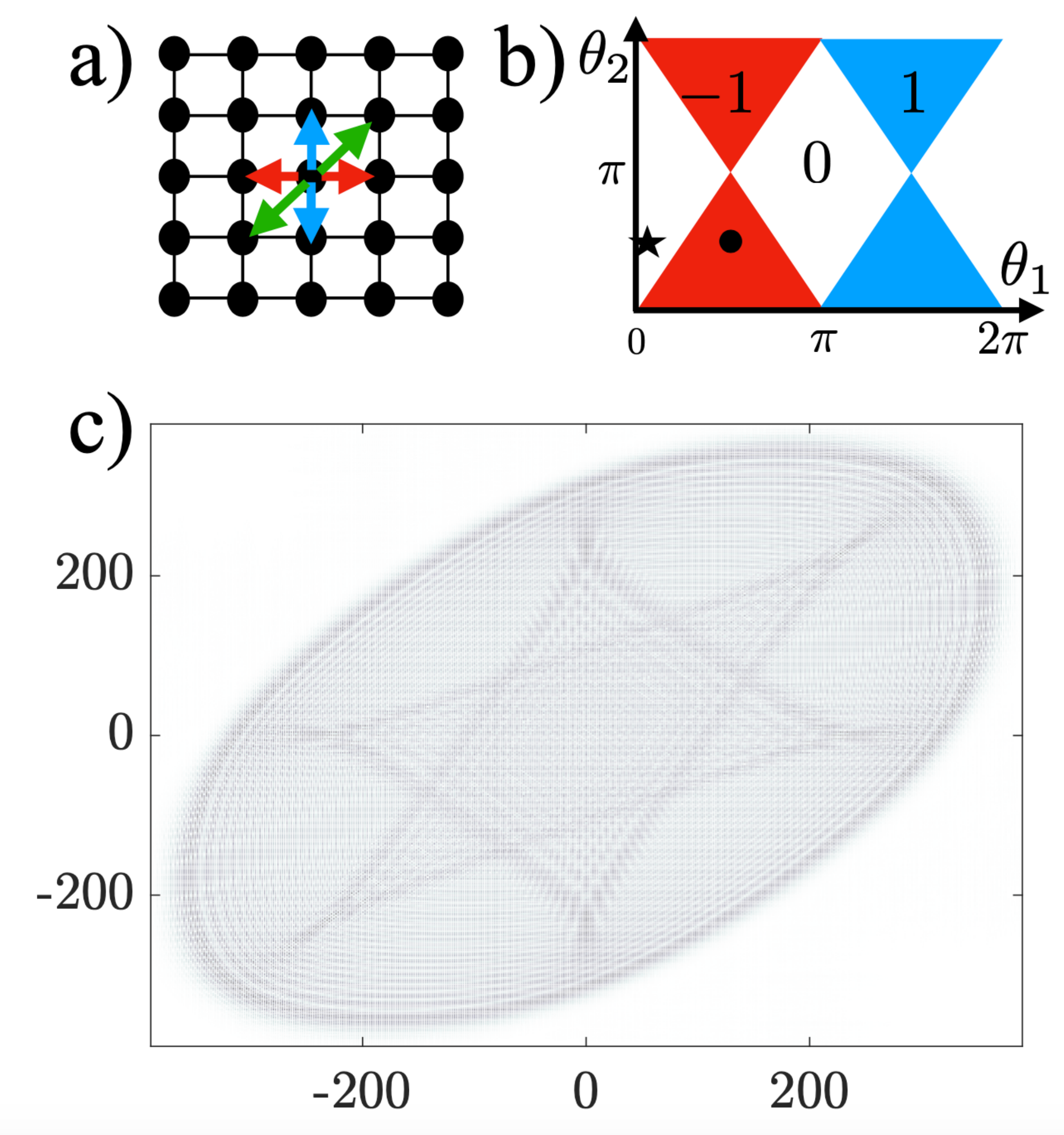}
    \caption{Topologically protected three-step quantum walk on a square lattice. a): The unitary of the split-step walk consists of three coin operations followed by translations in three different directions indicated by the arrows. b): Topological phase diagram of quantum walks as a function of the two coin parameters $\theta_1, \theta_2$. Colours correspond to different Chern numbers $C=0,1,-1$. c): Probability distribution of a topologically nontrivial walk $(\theta_1,\theta_2)=(\frac{\pi}{2},\frac{\pi}{2})$ at $t=256$. To enhance visibility, the colour scale varies linearly in $\|\psi\|^{1/2}$.}
    \label{fig:2}
\end{figure}

This unitary is parameterized by two coin angles, $\theta_{1}$ and $\theta_{2}$, and a single step of the walk consists of three sequential applications of combined spin flip and shift operations, as illustrated in Fig.~\ref{fig:2} a). 
The unitary \eqref{eq:dtqw_2d} can be  represented as $\hat{U}_{\mathrm{2D}}=e^{-\mathrm{i}\hat{H}_{\mathrm{eff}}}$, where the effective Hamiltonian in the momentum space is defined as
\begin{equation}
    \hat{H}_{\mathrm{eff}} = \int_{-\pi}^\pi d\vec{k} \Big[E(\mathbf{k})\mathbf{n}(\mathbf{k}) \cdot \hat{\sigma}\Big]\otimes \ket{\vec{k}}\bra{\vec{k}}.
\label{eq:h_eff}
\end{equation}

Here $\ket{\vec{k}}$ denotes the Fourier transformed position vector $\ket{\vec{r}}$. While the expressions for $E(\mathbf{k})$ and $\mathbf{n}(\mathbf{k})$, derived in Appendix A, are not particularly illuminating, the effective Hamiltonian can be analyzed with the tools of topological condensed matter theory to gain further insight on related quantum walks. In particular, the spectrum of the effective Hamiltonian is characterized by a topological index, the Chern number, which can be obtained by 
\begin{equation} \label{eq:chern}
    \mathcal{C} = \frac{1}{4\pi} \int_{\Omega} \mathbf{n}(\mathbf{k}) \cdot (\partial_{k_{x}} \mathbf{n}(\mathbf{k}) \times \partial_{k_{y}} \mathbf{n}(\mathbf{k})) \, \mathrm{d}\mathbf{k},
\end{equation}
where $\Omega$ denotes a torus $(k_{x}, k_{y}) \in \left[-\frac{\pi}{2}, \frac{\pi}{2}\right] \times \left[-\frac{\pi}{2}, \frac{\pi}{2}\right]$. In Fig.~\ref{fig:2} b) we have presented the topological phase diagram in terms of the coin parameters. A substantial fraction of the coin's parameter space supports non-zero Chern numbers, a primary motivation to study the walk protocol \eqref{eq:dtqw_2d}. 
As understood during the last four decades in condensed matter physics, topological states of matter are extraordinarily robust to disorder and Anderson localization. Electronic bands in integer quantum Hall systems, characterized by nonzero Chern numbers, are particularly striking examples of this. It is well-known by now that an arbitrarily weak randomness may lead to Anderson localization in 1d and 2d systems~\cite{RevModPhys.80.1355}. This means that with exponential accuracy, all the eigenstates of a Hamiltonian and the corresponding time-evolution unitary have a finite spatial support. Correspondingly, quantum walks defined by a unitary incorporating effects of disorder similarly give rise to walks that are exponentially confined to their initial position. However, as long as a system supports a nonzero Chern number, it is guaranteed to support at least some extended states \cite{RevModPhys.80.1355}. What this means for quantum walks is that, while protocols with trivial topology may enable extended walks in the presence of weak disorder, the topologically nontrivial protocols are always guaranteed to do so. Qualitatively, one expects topologically nontrivial systems to tolerate larger random perturbations before localizing. Since our main focus in this work is on random systems, it is natural to focus on topological split-step protocols. The propagation of a topological walk with finite Chern number on a square lattice is illustrated in Fig.~\ref{fig:2} c). In Appendix C, we also discuss two-step quantum walks which support topological Floquet-type phases. As shown below, these walks exhibit qualitatively similar properties on random lattices. 

\subsection{Split-step  walks on random lattices}

In the condensed-matter setting, it has been recognized for over half a century that any irregularities in a crystal lattice typically have dramatic effects on particle dynamics and transport properties. In particular, random disorder reduces a ballistic quantum propagation to diffusive motion. For quantum walks this implies that small fluctuations in the implementation of the quantum walk protocol could wipe out the quantum speed-up. This effect has also been studied in quantum walks by incorporating various sources of randomness in position and coin subspaces.~\cite{PhysRevA.74.022310,PhysRevLett.104.153602,kendon2007decoherence,PhysRevE.82.031122,PhysRevA.67.032304,PhysRevLett.106.180403,PhysRevA.98.032104,PhysRevA.96.033846,PhysRevB.98.134204}. Even worse, disorder can lead to the Anderson localization of all wave functions, so the spreading of the quantum walk could even stop altogether~\cite{konno2009localization,crespi2013anderson,PhysRevB.96.144204,PhysRevA.89.022309,PhysRevA.103.022416}. Thus, random irregularities could turn the quantum advantage of quantum walks into a quantum handicap. 

To study the fate of quantum walks on random geometries, we now generalize quantum walk protocols to random lattices. We consider an extensively-studied paradigm of random lattices, the site percolation on a square lattice~\cite{stauffer2018introduction}. The ensemble of percolation geometries is generated by assigning a probability $p$ for each site of a square lattice to be independently populated. Equivalently, each site is removed with probability $1-p$.  At low $p$, a single realization consists of a collection of disconnected clusters, whereas at high $p$ close to unity, the lattice resembles a square lattice with rare randomly missing sites. At the percolation threshold $p_{\text{perc}}\approx 0.59$, the system undergoes a geometric transition above which the system contains an infinite cluster connected by populated nearest-neighbour lattice sites~\cite{nakayama2003fractal}. A finite snapshot of a single realization of the $p=0.75$ ensemble is depicted in Fig.~1 a).

\begin{figure}[h]
    \centering
    \includegraphics[width=0.95\columnwidth]{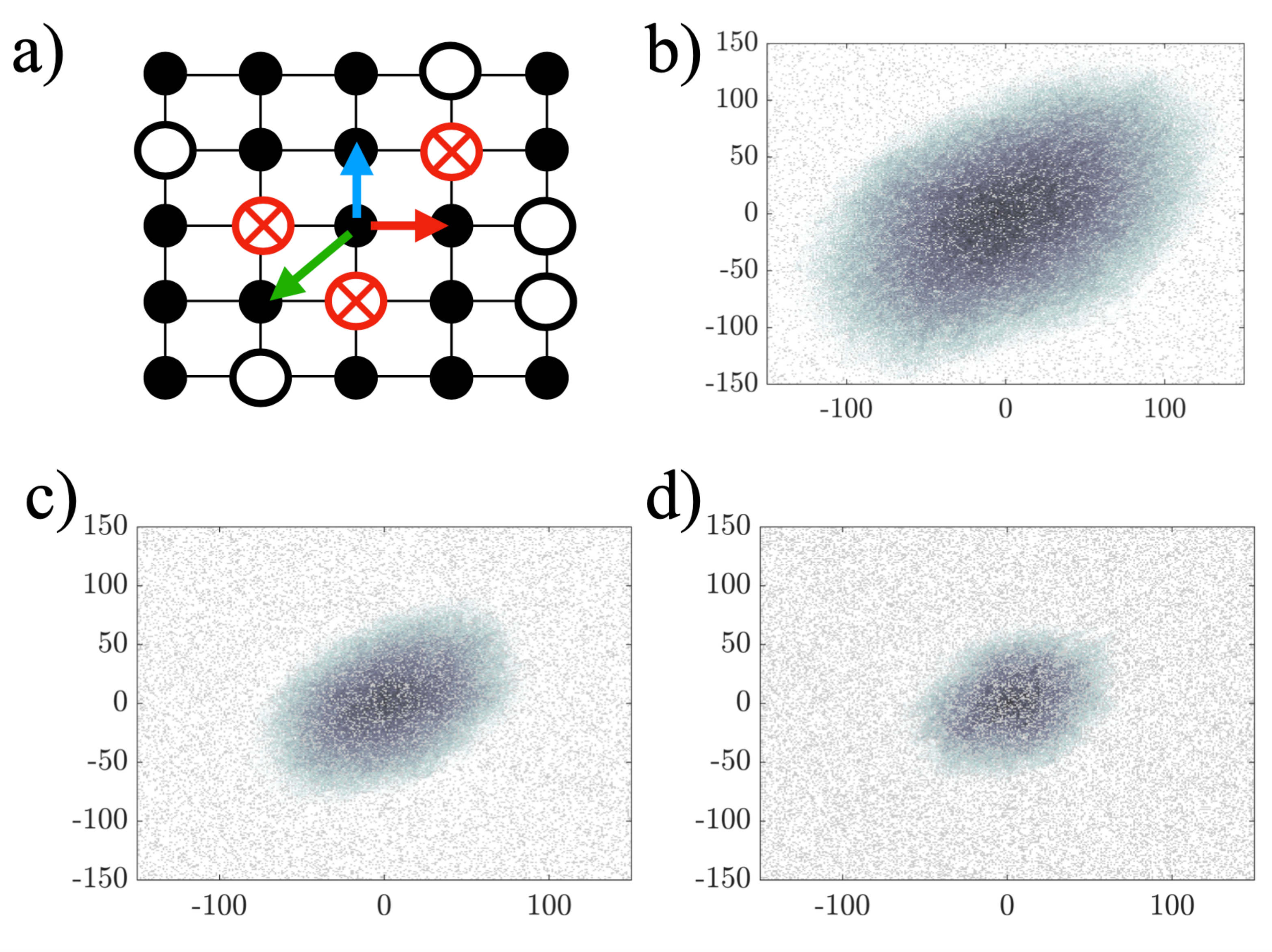}
    \caption{Quantum walks on lattices with randomly missing sites. a): Walker located in the center cannot hop on missing sites. The translations to prohibited sites, indicated by red color, are substituted by spin flips which reflect the walk to the opposite direction. b)-d): Probability distribution of quantum walks on a single realization of random lattices with $p=0.9$, $p=0.8$, $p=0.7$ after $t=256$ time steps, shown in logarithmic scale for improved visibility.}
    \label{fig:3}
\end{figure}
As illustrated in Fig.~\ref{fig:3} a), to generalize the protocol~\eqref{eq:dtqw_2d} to percolation lattices, it is necessary to modify translation operators when a walker is encountering absent sites. We implement a prescription where, instead of carrying out translation from site $\vec{r}$ to a missing site at $\vec{r}+\vec{\delta}$, the translation is omitted but the coin state of a walker is flipped to the opposite state $\ket{\vec{s}}\to\ket{\bar{\vec{s}}}$. This will prohibit the walker from entering the missing sites, effectively reflecting it to the opposite direction. A similar procedure has been employed in generating reflecting boundaries and boundary modes in topological quantum walks~\cite{kitagawa2012topological}. Thus, this prescription results in a random walk-generating unitary where the translation operator $\hat{T} $ is replaced by a spin-flip operation $\hat{S}$ when encountering randomly missing site. This construction, derived in detail in Appendix B, preserves the total probability associated with the wave function of the quantum walk. Examples of quantum walks for specific realizations of random lattices are illustrated in Figs.~\ref{fig:3} b)-d). The same prescription can be employed to generalize the two-step walk in Appendix C to lattices with missing sites.

Naturally, the unitary for a specific random lattice realization is not translation invariant and no longer admits a simple expression in momentum space. Nevertheless, since the generating unitary formally resembles a short-range hopping tight-binding Hamiltonian, the recent works on amorphous topological insulators~\cite{PhysRevResearch.2.013053,PhysRevResearch.2.043301} suggest that it is likely to maintain its topological character and avoid Anderson localization even on substantially diluted random lattices. Indeed, as seen below, the topologically nontrivial walks with a nonzero Chern number in the clean limit, are found to be quantitatively more robust to randomness compared to trivial walks with a vanishing Chern number.  

Being paradigmatic examples of random lattices, aspects of percolation-type geometries and quantum walks have been addressed in a number of previous works. These works have concentrated on 1d systems~\cite{rigovacca2016two}, finite graphs~\cite{kollar2014discrete}, time-dependent percolation configurations~\cite{PhysRevLett.108.230505,elster2015quantum,PhysRevLett.119.220503}, directed percolation~\cite{chandrashekar2014quantum} and continuous-time walks~\cite{chawla2019quantum}. To our knowledge, the only previous work attempting to analyze the asymptotic long-time dynamics on 2D percolation geometries is Ref.~\cite{leung2010coined}. As discussed below, this study was restricted to two orders of magnitude shorter time scales, which made it impossible to disentangle different kinematic regimes and the asymptotic phase diagram shown in Fig.~\ref{fig:1} c).

\section{Diffusion and localization on random lattices}

In this section we analyze statistical properties of quantum walks in ensembles of lattices corresponding to fixed random dilution, using fixed occupation probability $p$. To characterize the spreading speed, we define a MSD $\Delta^2X(t)=\overline{\bra{\psi(t)} X^2\ket{\psi(t)}}$, where the bar denotes a configuration average over different random lattice realizations with fixed $p$, and $\ket{\psi(t)}$ represents the quantum state of a walk which is located at the origin at $t=0$ with a specific coin state. Unless otherwise stated, the coin state at the origin corresponds to an eigenstate of $\sigma_y$ (different initial coin states are considered in Appendix D with similar conclusions). As suggested by the diffusive appearance of Figs.~\ref{fig:3} b)-d), we will see below that the long-time behavior of quantum walks is captured by the generalized diffusion Ansatz $\lim_{t\to\infty} \Delta^2X(t)\sim D_\alpha t^\alpha$, where $D_\alpha$ is a diffusion constant and $\alpha$ is a diffusion exponent. We focus on the two fundamental issues: i) What is the asymptotic propagation speed parametrized by $\alpha$? In particular, since a pristine lattice with $p=1$ supports a ballistic spreading with $\alpha=2$, is it possible to maintain a quantum speed-up with superdiffusive $\alpha>1$ for some degree of randomness $p<1$? ii) What is the critical dilution strength $0<p_c<1$ below which the walk localizes with $\alpha=0$? We furthermore explore how these issues are affected by topology of the walk protocol. 
\begin{figure}
    \centering
    \includegraphics[width=1.\columnwidth]{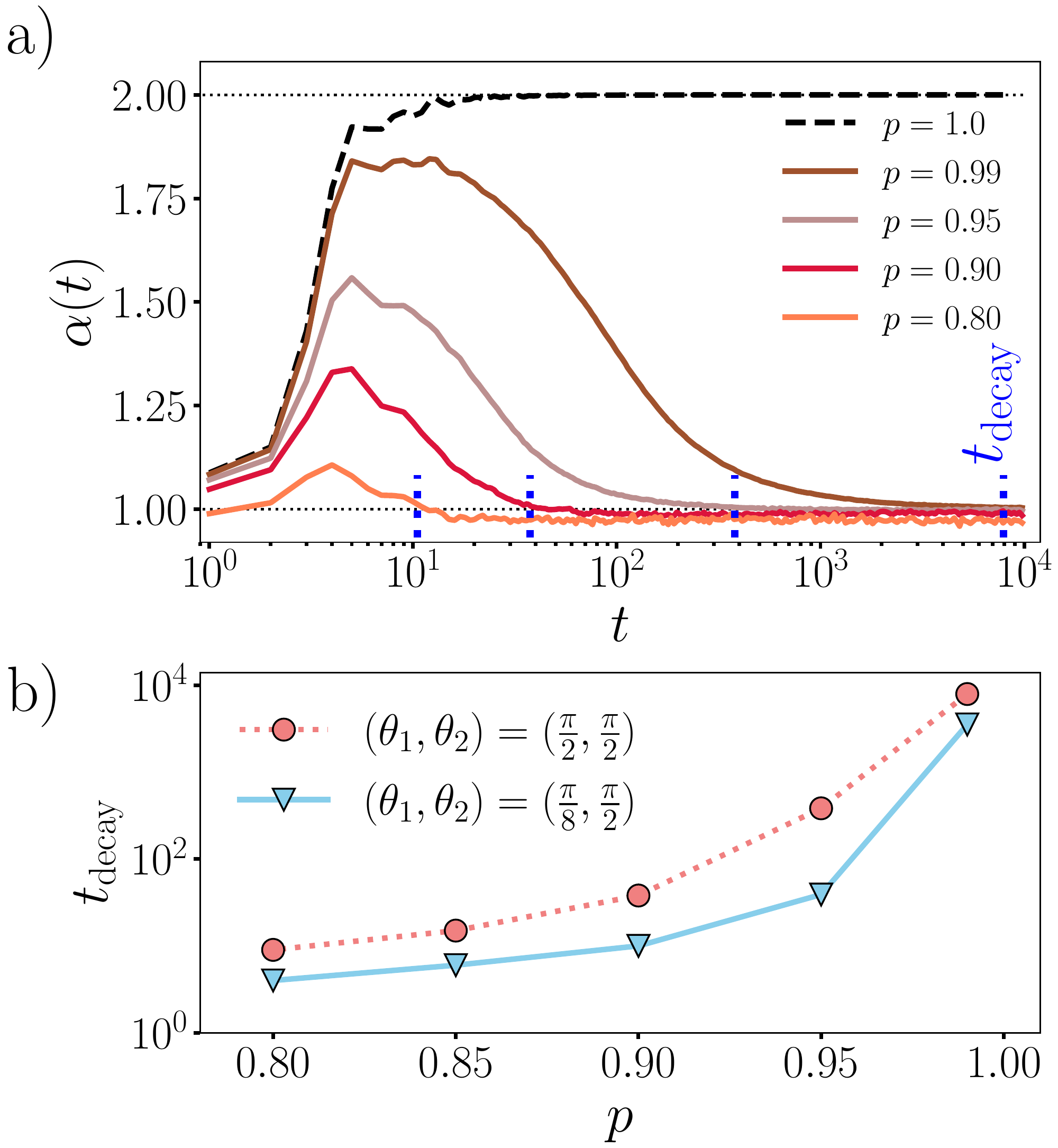}
    \caption{Determining the decay time (length) scale of the walker's propagation. The walker experiences a superdiffusive motion up to the time scale $t<t_{\text{decay}}$. a): Shown for different values of $p$ in the topological phase $\theta_1=\theta_2=\pi/2$. In the clean case $p=1$, the exponent quickly converges to the expected value of $\alpha(t\to\infty)=2$. b): This time scale depends on both the rate of the structural disorder $p$ and topological feature of the model controlled by the pair $(\theta_1,\theta_2)$. At $p=1$ the decay time is $t_{\text{decay}}=\infty$.
    }
    \label{fig:4}
\end{figure}

\begin{figure*}[ht]
    \centering
    \includegraphics[width=1.95\columnwidth]{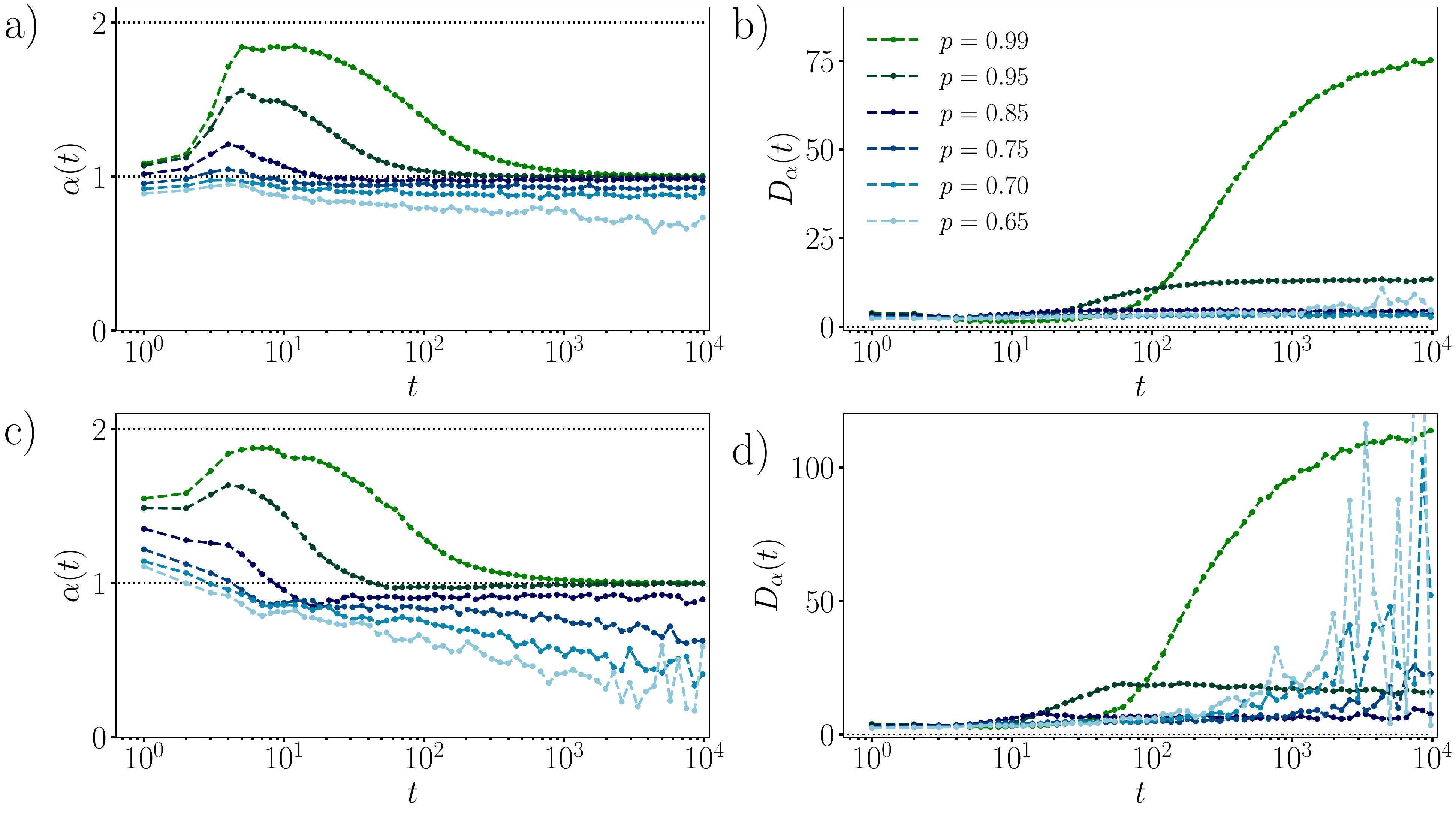}
    \caption{Diffusion parameters for configuration averaged DTQWs for various occupation probabilities $p$. a)-b): In the topological paramete space with $(\theta_1,\theta_2)=(\frac{\pi}{2},\frac{\pi}{2})$ and c)-d) trivial with  $(\theta_1,\theta_2)=(\frac{\pi}{8},\frac{\pi}{2})$ as a function of time steps. Fig.~a) and c) show that the long-time dynamics in delocalized regime is diffusive for all $p<1$. Near the localization threshold $p\approx 0.7$ the walk exhibits subdiffusion with $\alpha\approx 0.9$. The MSD data is averaged over $200-1000$ independent configurations at each time-step.
      }
    \label{fig:5}
\end{figure*}

We employ large-scale parallel computing to simulate quantum walks up to $t=10^4$ time steps and perform configuration averages up to $1000$ random lattice realizations. In general, smaller $p$ values require more configurations to damp the statistical fluctuations. The lattice size in simulations are effectively infinite; when the quantum walk approaches the specified computational boundary in some direction, the lattice size is increased to accommodate for the spreading so that the walker never encounters the boundaries. We analyze the quantum walk spreading in terms of the time-dependent function 
\begin{equation}\label{eq:alpha}
\alpha(t)=\frac{d\, \ln \Delta^2 X(t)}{d\,\ln t },
\end{equation}
which, for diffusive motion, reduces to the generalized diffusion exponent in the long-time limit  $\alpha=\lim_{t\to\infty}\alpha(t)$. However, by employing \eqref{eq:alpha}, we do not make any assumptions about the specific nature of the propagation. The qualitative behavior of $\alpha(t)$ at low dilution is presented in Fig.~\ref{fig:4} a) which illustrates that after an initial transient of a few time steps, $\alpha(t)$ achieves its maximum value. Remarkably, all but the clean system, which settles to ballistic motion $\alpha=2$, are seen to reduce to diffusive motion within some time scale $t_{\text{decay}}$. The decay time diverges as $p\to 1$, however, any finite randomness $p<1$ is seen to reduce the motion to diffusive as discussed below. In Fig.~\ref{fig:4} b) we compare the decay time of a topological and a trivial quantum walk indicated by a black circle and a star in Fig.~\ref{fig:2} b). For all $p$, the time scale $t_{\text{decay}}$ for the trivial walk with $(\theta_1,\theta_2)=(\frac{\pi}{8},\frac{\pi}{2})$ is shorter; that is, the trivial walks reach the ordinary diffusive steady state quicker. This is the first indication that the trivial walk is more susceptible to randomness and could be regarded as a precursor of the localization behavior discussed below.

To investigate diffusion and localization systematically, in Fig.~\ref{fig:5} we have illustrated the results for two quantum walk protocols which, in a clean system, correspond to finite and zero Chern numbers, respectively. We also plot the function $D_{\alpha}(t)=\Delta^2 X(t)/t^{\alpha(t)}$, which approaches the diffusion constant when $\alpha(t)$ converges to a non-zero constant. The coin parameters for these walks are indicated by the star and the circle in Fig.~\ref{fig:2} b). In the high-density regime, despite the initial superdiffusive transient $\alpha(t)>1$, even $1\%$ random dilution ($p=0.99$) leads to diffusive motion $\alpha=1$ in the long time limit. This indicates that, in the presence of random geometry, quantum walks \emph{cannot support  parametric quantum speed-up} compared to usual random walks. As seen in Fig.~\ref{fig:5} a),  $\alpha(t)$ for the topological walk converges to a constant when the density is higher than the threshold value $p>p_c\sim 0.70$. For densities $p<p_c$, the value of $\alpha(t)$ has a negative derivative which is consistent with localized quantum walk $\lim_{t\to\infty}\alpha(t)=0$. In contrast, as seen in Fig.~\ref{fig:5} c), the trivial walks localize earlier, around $p_c\sim 0.85$. Thus, the studied nontrivial walk can tolerate up to twice the density of randomly missing sites before localizing, confirming the general expectation that topological walks are more robust to random geometry. In Appendix D, we present results for diffusion quantities obtained for cases where $\alpha(t)$ converges to a constant, and we also plot results for quantum walks residing on the topological-trivial phase boundary, which indicate that their properties interpolate between the above-presented cases. The localization of quantum walks can be attributed to Anderson localization, a much-studied condensed-matter phenomenon that suppresses diffusion of quantum particles in random systems. By reducing $p$, percolation lattices also become geometrically disconnected, which could halt the quantum walks. However, on a square lattice, the nearest-neighbour disconnection happens only at lower density $p_{\text{perc}}\approx 0.59$, and is not relevant for the observed localization.

While it is challenging to pinpoint the exact localization threshold $p_c$ and the diffusion exponent at $p=p_c$, quantum walks near $p_c$ exhibit subdiffusive motion with $\alpha<1$. This behavior is clearly observed in Figs.~\ref{fig:5} a) and c), and is supported by the results in Appendix D. 
Our results indicate that for Chern walks the diffusion exponent drops down to $\alpha\lesssim 0.9$ before the system becomes localized. Qualitatively, the subdiffusive behavior can be regarded as a generic consequence of Anderson localization transitions~\cite{PhysRevLett.93.190503,PhysRevResearch.3.023052,PhysRevLett.100.094101,PhysRevLett.122.040501}. The eigenfunctions of the walk-generating unitary are qualitatively similar to a random tight-binding Hamiltonian which, at the localization threshold $p_c$, are known to become fractal. The anomalous diffusion is known to result from the interplay of diffusion and fractal geometry~\cite{nakayama2003fractal,RevModPhys.66.381,xu2021quantum,PhysRevA.32.3073}, both ingredients that are present at the localization threshold.

The above results constitute a radical departure from those of Ref.~\cite{leung2010coined}, which also studied diffusion of 2d quantum walks on percolation lattices, with the conclusion that the system supports continuous spectrum of different diffusion constant $0\leq\alpha\leq 2$ depending on $p$. However, this conclusion was reached by numerically propagating walks only up to relatively  short times of the order of $10^2$. As clearly seen in Fig.~\ref{fig:5}, the transient effects in studied walks are dominating on that time scale which prevents making definite conclusions regarding asymptotic motion. Obtaining a qualitatively correct picture of distinct dynamical regimes requires propagating walks for orders of magnitude longer times. Specifically, we see that a reliable picture of the pre-diffusive, diffusive and localized behavior in the studied system emerges only when analyzing walks up to $10^4$ time steps.

To further support the generic phase diagram in Fig.~\ref{fig:1} c) summarizing our results, in Appendix C we have studied a different quantum walk protocol and analyzed the qualitative features of the diffusive motion on 2d random lattices. This model also reproduces the key features observed above in the Chern walks---the absence of superdiffusion for any amount of randomness,  Anderson localization at $p_c$ above the geometric percolation threshold $p_{\text{perc}}$, and the subdiffusion near the localization threshold.

\section{Discussion}
 
Above we have seen how quantum walks on studied random lattices unavoidably lose their parametric quantum speed-up compared to ordinary diffusion. This discovery has dramatic implications for quantum information applications. Various efficient quantum-walk-based algorithms for the search of marked items in unstructured  databases and graphs invariably rely on the quantum speed-up. For example, in the pioneering work~\cite{PhysRevA.67.052307} it was shown that by combining quantum walks in hypercubic lattices with oracle queries, it is possible to achieve the quadratic quantum speed-up as in Grover's algorithm for the unstructured database search. Similarly, quantum-walk-based search of a 2d grid enjoys similar speed up with logarithmic correction~\cite{PhysRevA.70.022314,ambainis2004coins}. These speed-ups are enabled by the quadratic kinematic speed-up of quantum walks compared to random walks. As a natural generalization, one could ask whether quantum-walk-based spatial search algorithms on irregular lattices and random graphs could enjoy similar parametric speed-ups. For example, could quantum walks speed up the search for a marked item on a random maze generated by a percolation process, such as the one illustrated in Fig.~\ref{fig:1}(a)? Our results imply severe limitations to these hopes. 

The fact that the studied topologically-protected quantum walks always reduce to diffusion or subdiffusion in the long-time limit seems to offer little hope for achieving a parametric quantum speed-up in quantum-walk-based spatial search on generic random geometries. There are of course caveats: due to the unlimited possibilities in defining different walk protocols and random lattices, the present work cannot rule out that a special class of walks and graphs could escape the diffusive slowdown. However, it is instructive to assess this possibility through the lens of condensed matter physics, which has a long history of studying disordered quantum systems. In the condensed-matter setting, it has been widely observed that system-specific properties, such as microscopic details of disorder and its precise statistical properties are often irrelevant. The important factors for the qualitative behavior typically are discrete symmetries, spatial dimension and topological properties. In fact, if this was not the case, there would be little hope to model the behavior of realistic materials, for which the source of imperfections and their precise nature is often incompletely understood. This suggests that our finding of the asymptotic diffusive slowdown qualitatively applies to a much more general class of walks and random lattices and is not particular to the studied systems. Moreover, the experience from amorphous materials suggest that topological systems are exceptionally robust to the studied geometric randomness~\cite{PhysRevResearch.2.013053,PhysRevResearch.2.043301}. Thus, there are good reasons to believe that the studied topologically-protected walks are particularly robust examples of possible 2d quantum walks, and that the diffusive slowdown on random geometries is generic.  

Our results do not mean that quantum-walk-based algorithms on sufficiently small graphs could not outperform the speed of classical random walks. If one considers a finite snapshot of a percolation lattice, such as the one depicted in Fig.~\ref{fig:1} a), a quantum walk could very well require less steps than a random walk to effectively cover it. As seen in the previous section, quantum walks reduce to diffusive motion after a finite number of steps quantified by $t_{\text{decay}}$. This time scale also sets the linear size up to which quantum walks could be expected to maintain a possible advantage over diffusion. However, as seen in Fig.~\ref{fig:4}, the spreading of the walk begins to be affected already early on, even as soon as the walker bumps into the first missing site, which strongly reduces the speed of propagation. These factors seem to set a rather narrow window for quantum-walk-based advantages over random walks in random geometries.

\section{summary}

 In this work we studied 2d quantum walks on random lattices generated by site percolation processes. By carrying out large-scale simulations up to $10^4$ time steps, we identified a phase diagram of different diffusive regimes.  We observed that at long time scales even a small random dilution gives rise to a complete breakdown of the superdiffusive quantum speed-up, the hallmark of quantum walks on regular lattices. Additionally, we also identified a pre-diffusive regime, dominating at intermediate time scales, as the most promising setting for quantum walks to have an advantage over random walks.  Increasing the density of random defects will eventually halt quantum walks well above the geometric connectivity transition due to the Anderson localization. Near the localization threshold, the quantum walks exhibit sub\-diffusive spreading. Our observations imply stringent limitations for obtaining quantum speed-up in applications of quantum walks on random lattices and graphs.

\acknowledgements
The authors acknowledge the Academy of Finland for support.

\appendix

\section{Split-step walks on square lattice}
Here we outline the derivation of the effective Hamiltonian \eqref{eq:h_eff} for the square lattice quantum walk. The generating unitary for the studied three-step walk is 
\begin{align}
    \hat{U}_{\mathrm{2D}}(\theta_{1},\theta_{2}) = \hat{T}(\vec{\delta}_3)  \hat{R}(\theta_{1}) \hat{T}(\vec{\delta}_2) \hat{R}(\theta_{2})  \hat{T}(\vec{\delta}_1)  \hat{R}(\theta_{1}).
\label{eq:Uapp1}
\end{align}
where the coin operation is given by $\hat{R}(\theta) = e^{-\frac{1}{2}i\theta\sigma_{y}}$ and the translation operation is defined as 
\begin{align}
        \hat{T}({\delta}_i) = \sum_{\vec{r} \in \mathbb{Z}^{2}}\Big[ \ket{\uparrow}\bra{\uparrow} \otimes  \ket{\vec{r}+\vec{\delta}_i}\bra{\vec{r}} +\nonumber  \ket{\downarrow}\bra{\downarrow} \otimes  \ket{\vec{r}-\vec{\delta}_i}\bra{\vec{r}} \Big] 
\end{align}
for the three primitive translation vectors $\delta_i$ defined in the main text. By introducing momentum eigenstates $\ket{\vec{r}}=\int _{\Omega}\frac{d\vec{k}}{{(2\pi)^2}}e^{i\vec{k}\cdot \vec{r}}\ket{\vec{k}}$, the translation operator can be written as 
\begin{align}
        \hat{T}({\delta}_i) = \sum_{\vec{k}}\Big[ e^{i\vec{k}\cdot \delta_i\sigma_z}\otimes  \ket{\vec{k}}\bra{\vec{k}} \Big],
\end{align}
and the generating unitary becomes
\begin{align}
    \hat{U}_{\mathrm{2D}}(\theta_{1},\theta_{2})& =   \sum_{\vec{k}}e^{i\vec{k}\cdot \vec{\delta}_3\sigma_z}e^{-\frac{1}{2}i\theta_1\sigma_{y}} e^{i\vec{k}\cdot \vec{\delta}_2\sigma_z} e^{-\frac{1}{2}i\theta_2\sigma_{y}}\times \nonumber\\ &e^{i\vec{k}\cdot \vec{\delta}_1\sigma_z} e^{-\frac{1}{2}i\theta_1\sigma_{y}}\otimes  \ket{\vec{k}}\bra{\vec{k}}.
\label{eq:Uapp2}
\end{align}

The product of six Pauli matrix exponents can be combined into a single exponent through repeated use of the exponential multiplication rule \begin{equation}
    e^{i a (\mathbf{n} \cdot \hat{\sigma})} e^{ib (\mathbf{m} \cdot \hat{\sigma})} = e^{i c (\mathbf{p} \cdot \hat{\sigma})},\nonumber
\end{equation}
where
\begin{equation}
    \mathbf{p} = \frac{1}{\sin{c}} \left[ \mathbf{n}\sin{a}\cos{b} + \mathbf{m}\sin{b}\cos{a} - \left(\mathbf{n} \times \mathbf{m}\right) \sin{a}\sin{b} \right],\nonumber
\end{equation}
and
\begin{equation}
    \cos{c} = \cos{a}\cos{b} - \left(\mathbf{n} \cdot \mathbf{m}\right) \sin{a}\sin{b}.\nonumber
\end{equation}

This process, while technically straightforward, leads to cumbersome analytic expressions. The final results can be expressed as $\hat{U}_{\mathrm{2D}}(\theta_{1},\theta_{2})=e^{-\mathrm{i}\hat{H}_{\mathrm{eff}}}$, where we have defined an effective Hamiltonian as
\begin{equation}
    \hat{H}_{\mathrm{eff}} = \int_{-\pi}^\pi d\vec{k} \Big[E(\mathbf{k})\mathbf{n}(\mathbf{k}) \cdot \hat{\sigma}\Big]\otimes \ket{\vec{k}}\bra{\vec{k}}.
\label{eq:h_eff_app}
\end{equation}
The effective Hamiltonian is determined by the quasienergy $E(\vec{k})$ and the winding vector $\vec{n}(\vec{k})$ which have complicated expressions in terms of the momentum components $k_i$ and the coin parameters $\theta_i$. To write them down explicitly, we define the following shorthands for trigonometric functions
\begin{align}
 s^{k}_{a \cdot x + b \cdot y} &\equiv \sin(ak_{x} + bk_{y})\nonumber\\
 c^{k}_{a \cdot x + b \cdot y} &\equiv \cos(ak_{x} + bk_{y})\nonumber\\
 s^{\theta}_{\alpha \cdot 1 + \beta \cdot 2} &\equiv \sin\left( \frac{\alpha\theta_{1} + \beta\theta_{2}}{2} \right)\nonumber\\
c^{\theta}_{\alpha \cdot 1 + \beta \cdot 2} &\equiv \cos\left( \frac{\alpha\theta_{1} + \beta\theta_{2}}{2} \right) \nonumber
\end{align}
With this convention, the quasienergy is given by the relation 
\begin{align}
    \cos{E} = c^{\theta}_{2 \cdot 1}c^{\theta}_{2}c^{k}_{x}c^{k}_{x+2 \cdot y} - c^{\theta}_{2}s^{k}_{x}s^{k}_{x+2 \cdot y} - s^{\theta}_{2 \cdot 1}s^{\theta}_{2}\left(c^{k}_{x}\right)^{2}
\end{align}
and winding vector becomes
\begin{equation}
\begin{aligned}
    &\mathbf{n}(\vec{k}) =\\  &\frac{1}{\sin{E}} \begin{bmatrix} 
    s^{\theta}_{2}s^{k}_{x}c^{k}_{x} - s^{\theta}_{2 \cdot 1 + 2}s^{k}_{x}c^{k}_{y}c^{k}_{x+y} + s^{\theta}_{2 \cdot 1 - 2}s^{k}_{x}s^{k}_{y}s^{k}_{x+y} \\
    -s^{\theta}_{2}\left(s^{k}_{x}\right)^{2} - s^{\theta}_{2 \cdot 1 + 2}c^{k}_{x}c^{k}_{y}c^{k}_{x+y} + s^{\theta}_{2 \cdot 1 - 2}c^{k}_{x}s^{k}_{y}s^{k}_{x+y} \\
    c^{\theta}_{2}c^{k}_{x}s^{k}_{x + 2 \cdot y} + c^{\theta}_{2 \cdot 1 + 2}s^{k}_{x}c^{k}_{y}c^{k}_{x+y} - c^{\theta}_{2 \cdot 1 - 2}s^{k}_{x}s^{k}_{y}s^{k}_{x+y}
    \end{bmatrix}.
\end{aligned}
\end{equation}
The topological properties of the quantum walk are encoded in $\mathbf{n}(\vec{k})$ and can be revealed by evaluating the Chern number \eqref{eq:chern} as discussed in the main text.

\section{Generating unitary on random lattices}

In this appendix, we derive the position space representation of the quantum walk unitary on randomly diluted 2d lattices. We consider a finite $M \times N$ square lattice such that $\ket{x} \in \mathbb{R}^{M}$ and $\ket{y} \in \mathbb{R}^{N}$. With a two-level coin basis $\ket{s} \in \mathbb{C}^{2}$ associated to every lattice point, the basis states describing a quantum walk on a square lattice are given by  $\ket{\psi} = \ket{x} \otimes \ket{y} \otimes \ket{s} \in \mathbb{C}^{2MN}$. The propagator in this representation corresponds to a unitary matrix $\hat{U}_{\mathrm{2D}}(\theta_{1},\theta_{2}) \in \mathbb{C}^{2MN \times 2MN}$. We now introduce a random dilution by partitioning the lattice into two disjointed sets $\mathcal{L}$ and $\mathcal{P}$, where $\mathcal{L}$ corresponds to the set of lattice sites available for the quantum walk, while $\mathcal{P}$ corresponds to the set of randomly removed lattice sites. Now, we define shift operators $\hat{T}_{x} \in \mathbb{R}^{MN \times MN}$ and $\hat{T}_{y} \in \mathbb{R}^{MN \times MN}$ as
\begin{equation}
\begin{aligned}
    \hat{T}_{x}\ket{x}\otimes\ket{y} &= \ket{x+1}\otimes\ket{y} \\
    \hat{T}_{y}\ket{x}\otimes\ket{y} &= \ket{x}\otimes\ket{y+1},
\end{aligned}
\end{equation}
with $\hat{T}_{x}$ and $\hat{T}_{y}$ represented by a matrix with ones on the lower sub-diagonal
\begin{align}
    \hat{T}_{x} = \begin{bmatrix} 
    0 & 0 & 0 & \dots & 0 & 0 \\
    1 & 0 & 0 & \dots & 0 & 0 \\
    0 & 1 & 0 & \dots & 0 & 0 \\
    \vdots & \vdots & \ddots & \vdots & \vdots \\
    0 & 0 & 0 & \dots & 1 & 0
    \end{bmatrix} \in \mathbb{R}^{M \times M}, \nonumber \\
    \hat{T}_{y} = \begin{bmatrix} 
    0 & 0 & 0 & \dots & 0 & 0 \\
    1 & 0 & 0 & \dots & 0 & 0 \\
    0 & 1 & 0 & \dots & 0 & 0 \\
    \vdots & \vdots & \ddots & \vdots & \vdots \\
    0 & 0 & 0 & \dots & 1 & 0
    \end{bmatrix} \in \mathbb{R}^{N \times N},
\end{align}
where periodic boundary conditions could be incorporated in $x$ and $y$ directions by setting the matrix elements $[\hat{T}_{x}]_{M-1,0}$ and $[\hat{T}_{y}]_{N-1,0}$ to 1.  Additionally, we define identity operators $\hat{I}_{x} \in \mathbb{R}^{MN \times MN}$ and $\hat{I}_{y} \in \mathbb{R}^{MN \times MN}$ as
\begin{equation}
\begin{aligned}
    \hat{I}_{x}\ket{x}\otimes\ket{y} &= \ket{x}\otimes\ket{y} \\
    \hat{I}_{y}\ket{x}\otimes\ket{y} &= \ket{x}\otimes\ket{y},
\end{aligned}
\end{equation}
which are identity matrices of appropriate dimensions. Finally, we define the lattice operator $\hat{L} \in \mathbb{R}^{MN \times MN}$ and percolation operator $\hat{P} \in \mathbb{R}^{MN \times MN}$. These operators act as the projectors to the two sublattices
\begin{align}
    \hat{L} \big[\ket{x} \otimes \ket{y}\big] = \begin{cases}
    1, \quad \text{if} \quad (x,y) \in \mathcal{L} \\
    0, \quad \text{if} \quad (x,y) \notin \mathcal{L} \\
    \end{cases} \nonumber \\
    \hat{P} \big[\ket{x} \otimes \ket{y}\big] = \begin{cases}
    1, \quad \text{if} \quad (x,y) \in \mathcal{P} \\
    0, \quad \text{if} \quad (x,y) \notin \mathcal{P} \\
    \end{cases},
\end{align}
with the corresponding matrix representations
\begin{equation}
\begin{aligned}
    \hat{L} &= \sum_{(x,y) \in \mathcal{L}} \big(\ket{x} \otimes \ket{y}\big) \big(\ket{x} \otimes \ket{y}\big)^{T} \\
    \hat{P} &= \sum_{(x,y) \in \mathcal{P}} \big(\ket{x} \otimes \ket{y}\big) \big(\ket{x} \otimes \ket{y}\big)^{T}.
\end{aligned}
\end{equation}
With all the relevant definitions in place, translation operators $\hat{T}_{1}, \hat{T}_{2}$ and $\hat{T}_{3}$ for the topological split-step walk on a diluted lattice can be written in the block matrix form:
\begin{widetext}
\begin{equation}
\begin{aligned}
    \hat{T}_{1} &= \begin{bmatrix} 
    \hat{L} (\hat{T}_{x} \otimes \hat{T}_{y}) & (\hat{T}_{x} \otimes \hat{T}_{y}) \hat{P} (\hat{T}_{x}^{\dagger} \otimes \hat{T}_{y}^{\dagger}) \\
    (\hat{T}_{x}^{\dagger} \otimes \hat{T}_{y}^{\dagger}) \hat{P} (\hat{T}_{x} \otimes \hat{T}_{y}) & \hat{L} (\hat{T}_{x}^{\dagger} \otimes \hat{T}_{y}^{\dagger})
    \end{bmatrix} \in \mathbb{R}^{2MN \times 2MN} \\
    \hat{T}_{2} &= \begin{bmatrix} 
    \hat{L} (\hat{I}_{x} \otimes \hat{T}_{y}) & (\hat{I}_{x} \otimes \hat{T}_{y}) \hat{P} (\hat{I}_{x} \otimes \hat{T}_{y}^{\dagger}) \\
    (\hat{I}_{x} \otimes \hat{T}_{y}^{\dagger}) \hat{P} (\hat{I}_{x} \otimes \hat{T}_{y}) & \hat{L} (\hat{I}_{x} \otimes \hat{T}_{y}^{\dagger})
    \end{bmatrix} \in \mathbb{R}^{2MN \times 2MN} \\
    \hat{T}_{3} &= \begin{bmatrix} 
    \hat{L} (\hat{T}_{x} \otimes \hat{I}_{y}) & (\hat{T}_{x} \otimes \hat{I}_{y}) \hat{P} (\hat{T}_{x}^{\dagger} \otimes \hat{I}_{y}) \\
    (\hat{T}_{x}^{\dagger} \otimes \hat{I}_{y}) \hat{P} (\hat{T}_{x} \otimes \hat{I}_{y}) & \hat{L} (\hat{T}_{x}^{\dagger} \otimes \hat{I}_{y})
    \end{bmatrix} \in \mathbb{R}^{2MN \times 2MN}.
\end{aligned}
\end{equation}
\end{widetext}
The block structure reflects the coin degree of freedom, so that the diagonal blocks in this matrix correspond to the standard translation of the walker when the hopping site is present ($\ket{\uparrow}\bra{\uparrow}$ and $\ket{\downarrow}\bra{\downarrow}$ terms). The off-diagonal blocks enforce the forbidden hopping to the missing sites as well as incorporate the reversal of the coin state ($\ket{\uparrow} \to\bra{\downarrow}$ and $\ket{\downarrow} \to \bra{\uparrow}$). Similarly, the coin operator $\hat{R}_{y}(\theta)$ has a block matrix representation
\begin{equation}
    \hat{R}_{y}(\theta) = \begin{bmatrix} 
    \cos{\frac{\theta}{2}} (\hat{I}_{x} \otimes \hat{I}_{y}) & \sin{\frac{\theta}{2}} (\hat{I}_{x} \otimes \hat{I}_{y}) \\
    -\sin{\frac{\theta}{2}} (\hat{I}_{x} \otimes \hat{I}_{y}) & \cos{\frac{\theta}{2}} (\hat{I}_{x} \otimes \hat{I}_{y})
    \end{bmatrix} \in \mathbb{R}^{2MN \times 2MN}.
\end{equation}
We then obtain the propagator by straightforward matrix multiplication
\begin{equation}
    \hat{U}_{\mathrm{2D}}(\theta_{1},\theta_{2}) = \hat{T}_{3}\hat{R}_{y}(\theta_{1})\hat{T}_{2}\hat{R}_{y}(\theta_{2})\hat{T}_{1}\hat{R}_{y}(\theta_{1}).
\end{equation}

\begin{figure}
    \centering
    \includegraphics[width=1.\columnwidth]{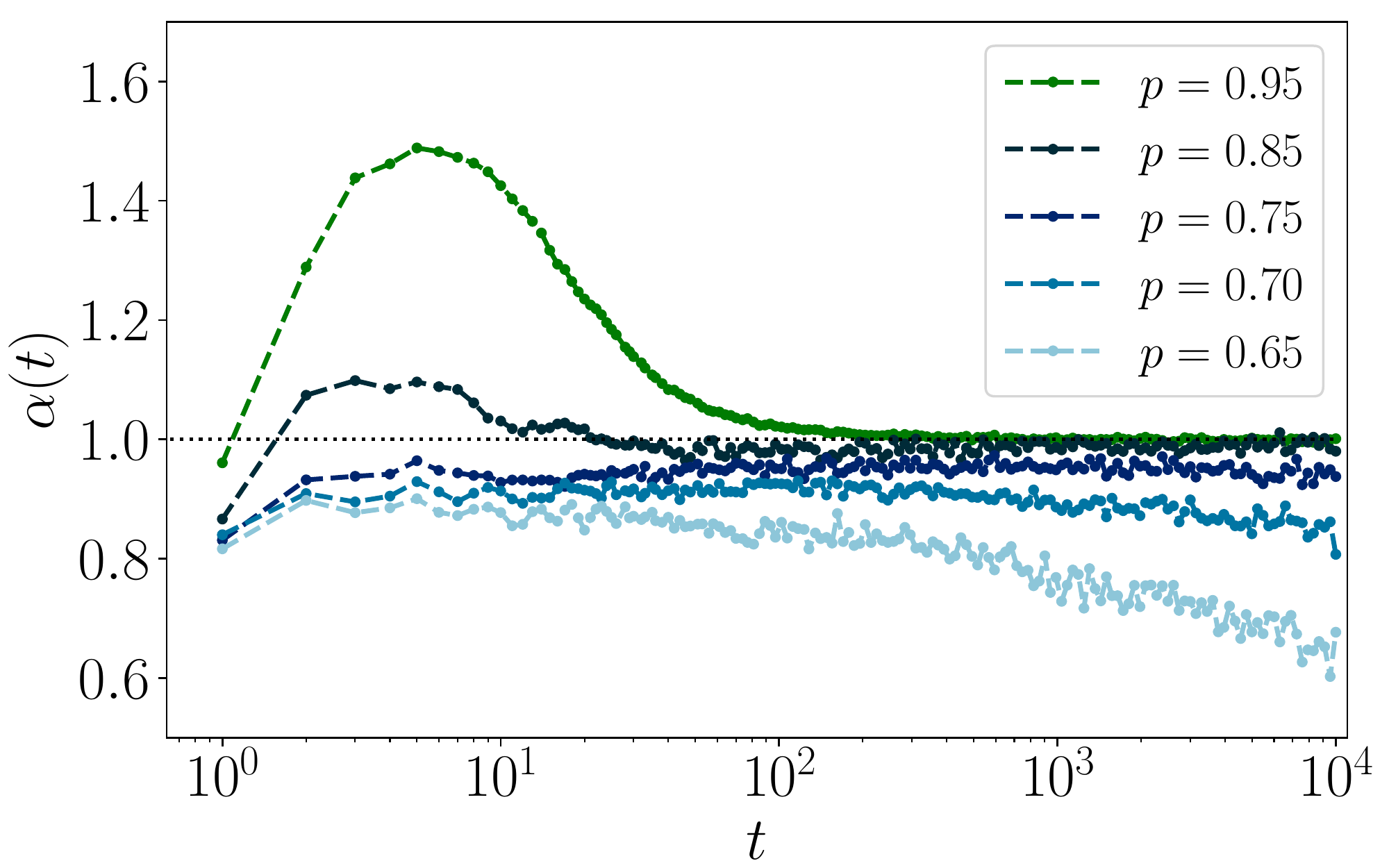}
    \caption{Evolution of the diffusion exponent at different $p$ for Eq. \ref{eq:dtqw_2d_app} with $(\theta_1,\theta_2)=(\frac{\pi}{2},0)$. The diffusion exponent is calculated using Eq.~\eqref{eq:alpha}, where the MSD data is averaged over 300-1000 independent configurations at each point.}
    \label{fig:2_appendix}
\end{figure}

\section{Topological quantum walks with Floquet invariants }
Here we consider a simpler walk where the diagonal translation $\hat{T}(\vec{\delta}_1)$ and the corresponding rotation operator $\hat{R}_{y}(\theta_{2})$ are dismissed:
\begin{align}
    \hat{U}_{\mathrm{2D}}(\theta_{1},\theta_{2}) = \hat{T}(\vec{\delta}_3)  \hat{R}(\theta_{1}) \hat{T}(\vec{\delta}_2) \hat{R}(\theta_{2}),
\label{eq:dtqw_2d_app}
\end{align}
following the same convention as introduced in the main text. This family of walks~\cite{kitagawa2012topological} has been realized experimentally~\cite{PhysRevLett.121.100502,PhysRevLett.121.100501,PhysRevLett.129.046401} and is characterized by the Rudner winding number~\cite{PhysRevX.3.031005}:
\begin{align}
    W[U]=&\frac{1}{8\pi^2}\int dt\,dk_x\,dk_y\nonumber\\
    &\times \text{Tr}(U^{-1}\partial_t\,U[U^{-1}\,\partial_{k_x}\,U,U^{-1}\,\partial_{k_y}\,U]).
\end{align}
In Fig.~\ref{fig:2_appendix}, we have plotted the dynamical exponent at $(\theta_1,\theta_2)=(\frac{\pi}{2},0)$ for various occupation probabilities $p$. In this parameter regime, the Floquet operator~\ref{eq:dtqw_2d_app} admits a finite winding number. Qualitatively, such walks exhibit the same response to the structural disorder, which generalizes the findings of the main text to quantum walks in different topological symmetry classes.

\section{Additional details on topological Chern walks }
\begin{figure*}[hb]
    \centering
    \includegraphics[width=1.8\columnwidth]{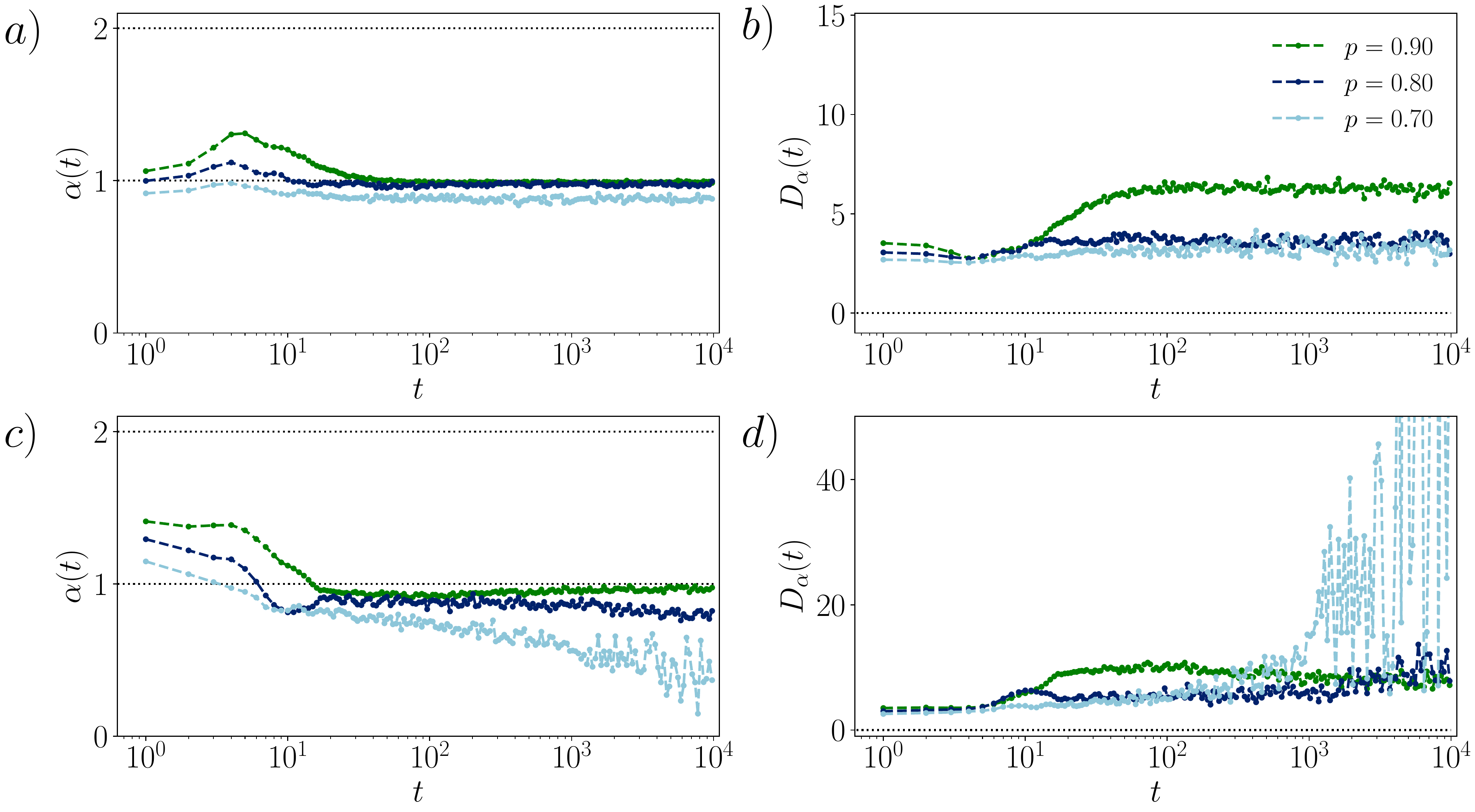}
    \caption{The evolution dynamical exponent $\alpha(t)$ and the diffusion constant $D_{\alpha}(t)$ at different occupation probabilities $p$. Here the walker is initialized in an eigenstate of $\sigma_x$. a)-b): In topological phase with $(\theta_1,\theta_2)=(\frac{\pi}{2},\frac{\pi}{2})$ and c)-d) trivial phase with $(\theta_1,\theta_2)=(\frac{\pi}{8},\frac{\pi}{2})$. Data is averaged over 500-700 random configurations. }
    \label{fig:3_appendix}
\end{figure*}

To establish the generic nature of the quantum walk on random lattices, it is important to exclude any initial state dependence effects. Since the operation on internal degrees of freedom is a rotation with respect to the $y$ axis, it is instructive to show that the behavior is invariant under initial states other than the one studied in the main text. In this way, we initialize the walker in an eigenstate of the Pauli spin operator $\sigma_x$ given by $\frac{1}{\sqrt{2}} [1,1]$. Fig.~\ref{fig:3_appendix} illustrates the response of the walker to various strength of the structural disorder starting from such an initial state. As can be seen, the key results reported in the main text such as asymptotic diffusive (classical) behavior, the existence of a finite region of sudiffusive spread and the eventual localization, all hold in this case as well.

\begin{figure*}[ht]
    \centering
    \includegraphics[width=1.8\columnwidth]{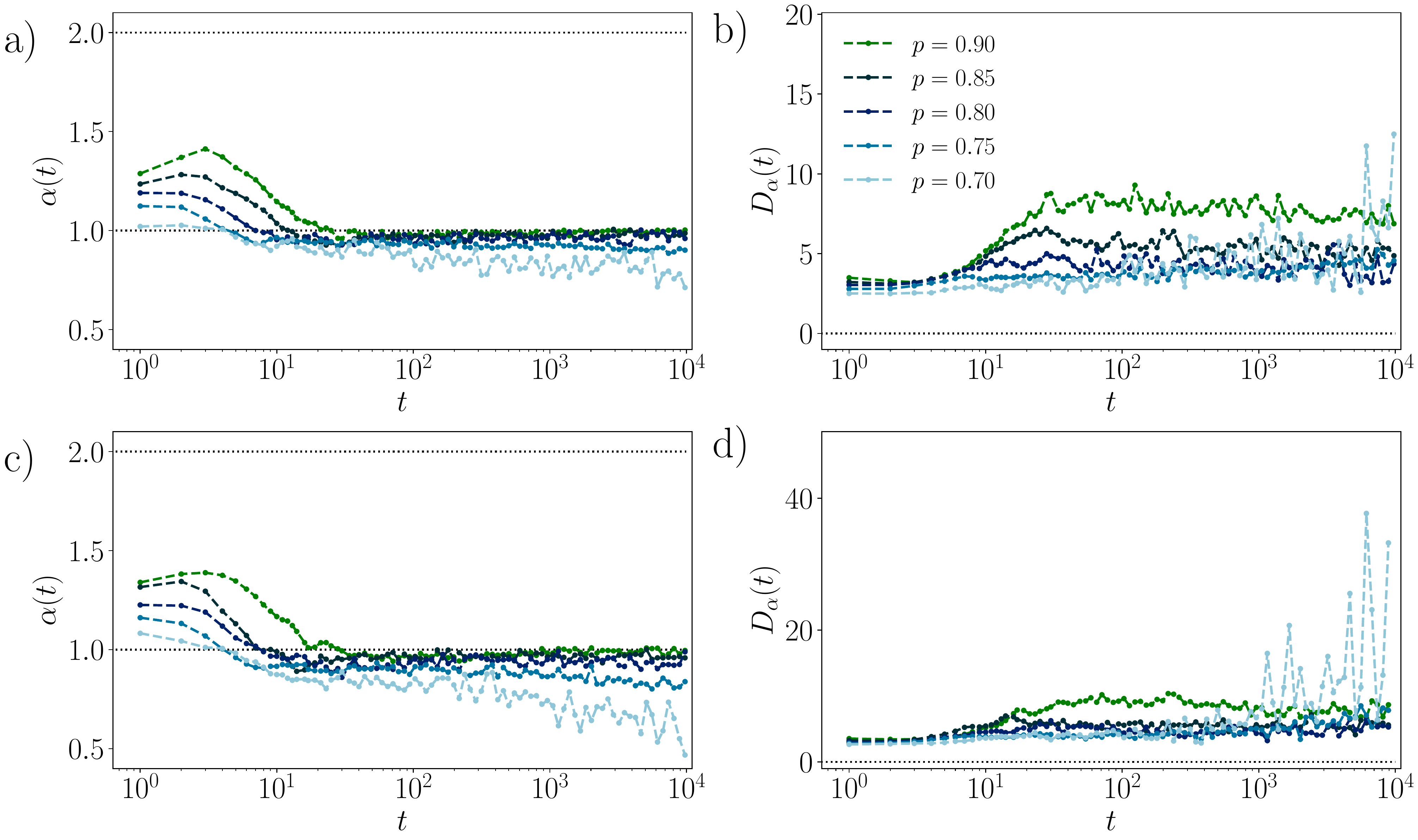}
    \caption{Dynamical exponent $\alpha(t)$ and the diffusion constant $D_{\alpha}(t)$ at different occupation probabilities $p$. a)-b): For $(\theta_1,\theta_2)=(\frac{\pi}{3},\frac{\pi}{2})$ and c)-d) $(\frac{\pi}{4},\frac{\pi}{2})$. }
    \label{fig:4_appendix}
\end{figure*}

Additionally, we consider the behavior of the walker at different points on the phase diagram (of the clean case) close and at the boundary of the topological-trivial transition. The results are shown in Fig.~\ref{fig:4_appendix}, hinting toward an intermediate behavior between the cases presented in the main text.

Lastly, for high enough occupation probability $p$, the diffusion exponent $\alpha(t)$ converges to some value, at least within the range of $10^4$ time steps. By fitting a line through the fluctuating curve using a least squares method, we can find the height of the horizontal line, corresponding to $\alpha(t \to \infty)$. For lower values of $p$, below the localization threshold, the diffusion exponent does not converge, and the above method is not applicable.
Additionally, through the generalized diffusion Ansatz $\lim_{t\to\infty} \Delta^2X(t)\sim D_\alpha t^\alpha$, we find the diffusion constant $D_\alpha$. Where applicable, these values are given in Tables~\ref{table:1} and \ref{table:2}.


\begin{table}[h]
\centering
\begin{tabular}{| c | c | c |}
  \hline
 $p$ & $\alpha(t\to\infty)$ & $D_\alpha(t\to\infty)$ \\ [1.5ex] 
 \hline\hline
 1.0  & 2.0  & 1.25 \\ 
 0.99 & 1.01 & 70.5 \\ 
 0.95 & 1.0  & 13.1 \\ 
 0.85 & 0.98 & 4.49 \\ 
 0.75 & 0.93 & 3.65 \\ 
 0.70 & 0.88 & 3.33 \\ 
 [1ex] 
 \hline
\end{tabular}
\caption{Fits for long time behavior of the diffusion exponent $\alpha(t \to \infty)$ and the diffusion exponent $D_\alpha$ for the generating unitary (\ref{eq:dtqw_2d}) in topological parameter regime $(\theta_1,\theta_2)=(\frac{\pi}{2},\frac{\pi}{2})$.}
\label{table:1}
\end{table}

\begin{table}[h]
\centering
\begin{tabular}{| c | c | c |}
  \hline
  $p$ & $\alpha(t\to\infty)$ & $D_\alpha(t\to\infty)$ \\ [1.5ex] 
 \hline\hline
 1.0  & 2.0  & 2.34 \\ 
 0.99 & 1.01 &  108 \\ 
 0.95 & 0.99 & 16.1 \\ 
 0.85 & 0.91 &  6.2 \\ 
 [1ex] 
 \hline
\end{tabular}
\caption{Fits for long time behavior of the diffusion exponent $\alpha(t \to \infty)$ and the diffusion exponent $D_\alpha$ for the generating unitary (\ref{eq:dtqw_2d}) in trivial parameter regime $(\theta_1,\theta_2)=(\frac{\pi}{8},\frac{\pi}{2})$.}
\label{table:2}
\end{table}

\clearpage
\bibliography{tqrw}

\end{document}